\documentclass{article}
\usepackage{array,hhline}
\usepackage{spconf,amsmath,graphicx}
\usepackage{stfloats}
\usepackage{float}
\usepackage[colorlinks=true,linkcolor=black,citecolor=blue,urlcolor=blue,]{hyperref}
\usepackage[doipre={doi:~}]{uri}

\usepackage{amssymb}
\usepackage{multirow}
\usepackage{booktabs}
\usepackage[misc]{ifsym}

\title{MIXED TRANSFORMER U-NET FOR MEDICAL IMAGE SEGMENTATION}
\name{\normalsize{Hongyi Wang$^1$
Shiao Xie$^1$
*Lanfen Lin$^1$
Yutaro Iwamoto$^2$
Xian-Hua Han$^3$
*Yen-Wei Chen$^{3,4,1}$
Ruofeng Tong$^{1,4}$}%
\thanks{Hongyi Wang and Shiao Xie contributed equally to this work.}}
\address{$^1$College of Computer Science and Technology, Zhejiang University, China\\
$^2$College of Information Science and Engineering, Ritsumeikan University, Japan\\
$^3$Artificial Intelligence Research Center, Yamaguchi University, Japan\\
$^4$Research Center for Healthcare Data Science, Zhejiang Lab, China}
\begin{document}

\maketitle
\begin{abstract}
Though U-Net has achieved tremendous success in medical image segmentation tasks, it lacks the ability to explicitly model long-range dependencies. Therefore, Vision Transformers have emerged as alternative segmentation structures recently, for their innate ability of capturing long-range correlations through Self-Attention (SA). However, Transformers usually rely on large-scale pre-training and have high computational complexity. Furthermore, SA can only model self-affinities within a single sample, ignoring the potential correlations of the overall dataset. To address these problems, we propose a novel Transformer module named Mixed Transformer Module (MTM) for simultaneous inter- and intra- affinities learning. MTM first calculates self-affinities efficiently through our well-designed Local-Global Gaussian-Weighted Self-Attention (LGG-SA). Then, it mines inter-connections between data samples through External Attention (EA). By using MTM, we construct a U-shaped model named Mixed Transformer U-Net (MT-UNet) for accurate medical image segmentation. We test our method on two different public datasets, and the experimental results show that the proposed method achieves better performance over other state-of-the-art methods. The code is available at: https://github.com/Dootmaan/MT-UNet.
\end{abstract}
\section{Introduction}
\label{sec:intro}

Automatic accurate medical image segmentation is of great significance for disease diagnosis nowadays. U-Net \cite{unet}, which consists of an encoder-decoder network with skip-connections, has been proved to be effective for many different segmentation tasks. Despite its dominant position in medical image processing, U-Net and its variants \cite{unet++,unet3+,attentionunet} also suffer from the problem that all the CNNs face: the lack of ability to model long-range correlations. This is mainly because of the intrinsic locality of convolution operations.

Recently, many works try to solve this problem by using Transformer encoder \cite{ViT, detr, transunet}. Transformer is an attention-based model originally designed for sequence-to-sequence prediction \cite{attentionisallyouneed}. Self-Attention (SA) is the key component of Transformer. It can model correlations among all the input tokens, giving Transformer the ability to handle long-range dependencies. Though some of these work achieved satisfying results \cite{transunet,transfuse,transbts,transclaw}, they usually rely heavily on large-scale pre-training, causing inconvenience to the use of the methods. In addition, SA has a quadratic computational complexity, which may slow down the processing speed for high-dimensional data such as medical images. Last but not least, SA also has the limitation of ignoring inter-sample correlations, leaving a large room for further improvements.

To tackle these problems, we redesign SA for better local perception with lower computational cost, then integrating it with External Attention (EA) \cite{externalattention} to manage inter- and intra-correlations simultaneously. Since in most vision tasks the visual dependencies between regions nearby are usually stronger than those far away, we perform local SA on fine-grained local context and global SA only on coarse-grained global context. When calculating global attention maps, we use Axial Attention \cite{axialattention} to reduce the amount of calculation, and further introduce a learnable Gaussian matrix \cite{GaussianTrans} to enhance the weight of nearby tokens. The main reason of Transformer requiring large-scale pre-training lies in that it has no prior knowledge about the structure of the problem. So when we design MT-UNet, we use Convolution Stem as feature extractor for the shallow layers, setting structure priors for the segmentation task. Experiments show that our method can surpass other state-of-the-art methods without pre-training.

In general, our contributions are three-folded: (1) We design MTM for simultaneous inter- and intra-affinities learning. (2) We propose LGG-SA, which perform SA sequentially on fine-grained local context and coarse-grained global context. We also introduce a learnable Gaussian matrix to emphasize the nearby areas around each query. (3) We build a Mixed Transformer U-Net for medical image segmentation and verify its effectiveness with two different datasets.

\section{Methods}
\label{sec:format}

\begin{figure}[t]

\centering
\includegraphics[width=0.48\textwidth]{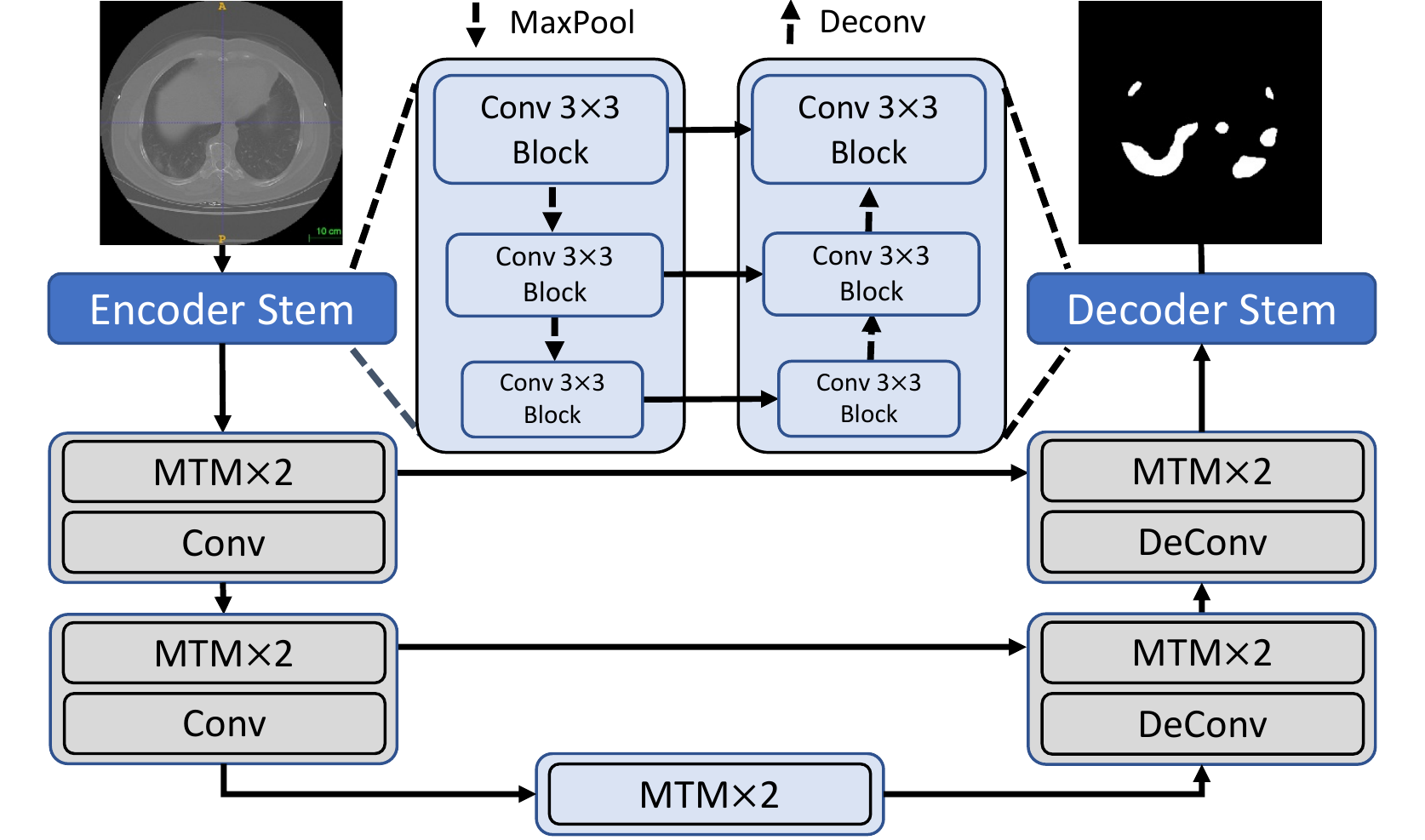}
\caption{A schematic view of the proposed MT-UNet. }
\label{fig1}
\end{figure}

\subsection{Overall Structure Design}
A schematic view of the proposed method is shown in Fig.~\ref{fig1}. The network is based on an encoder-decoder structure, and it uses skip connections to keep low-level features when decoding. As is shown, MTMs are only used for deeper layers with smaller spatial size to reduce the computational cost, while the upper layers still use classic convolutional operation. This is because we want to focus on local relations on the initial layers since they contain more high-resolution details. By using convolution, we can also introduce some structure priors to the model, which can be helpful for medical image datasets with relatively small size. It should be noted that for all the Transformer modules, a 2-stride convolutional/deconvolutional kernel is followed to realize down-sampling/up-sampling as well as channel expanding/squeezing.

\begin{figure}[t]
\centering
\includegraphics[width=0.48\textwidth]{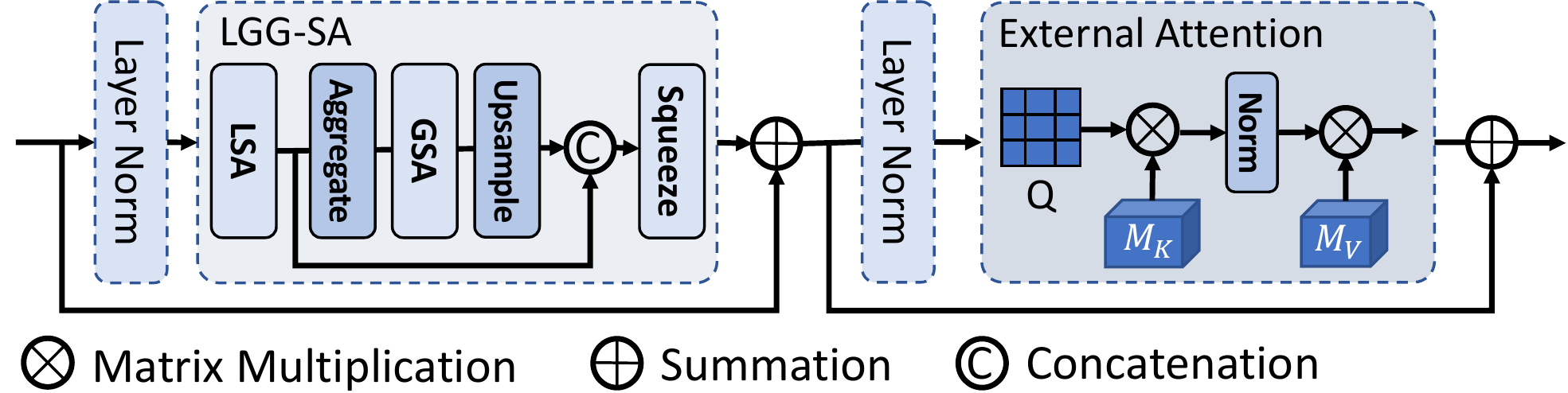}
\caption{Overview of the proposed Mixed Transformer Module.}
\label{fig2}
\end{figure}

\subsection{Mixed Transformer Module}
Overview of the proposed MTM is shown in Fig.~\ref{fig2}. As is presented, MTM consists of LGG-SA and EA. LGG-SA is designed to model short- and long-range dependencies with different granularity, while EA is used to exploit inter-sample correlations. This module is proposed to replace the original Transformer encoder for its better performance on vision tasks and lower time complexity.

\subsection{Local-Global Gaussian-Weighted Self-Attention}
\begin{figure*}[t]
\centering
\includegraphics[width=\textwidth]{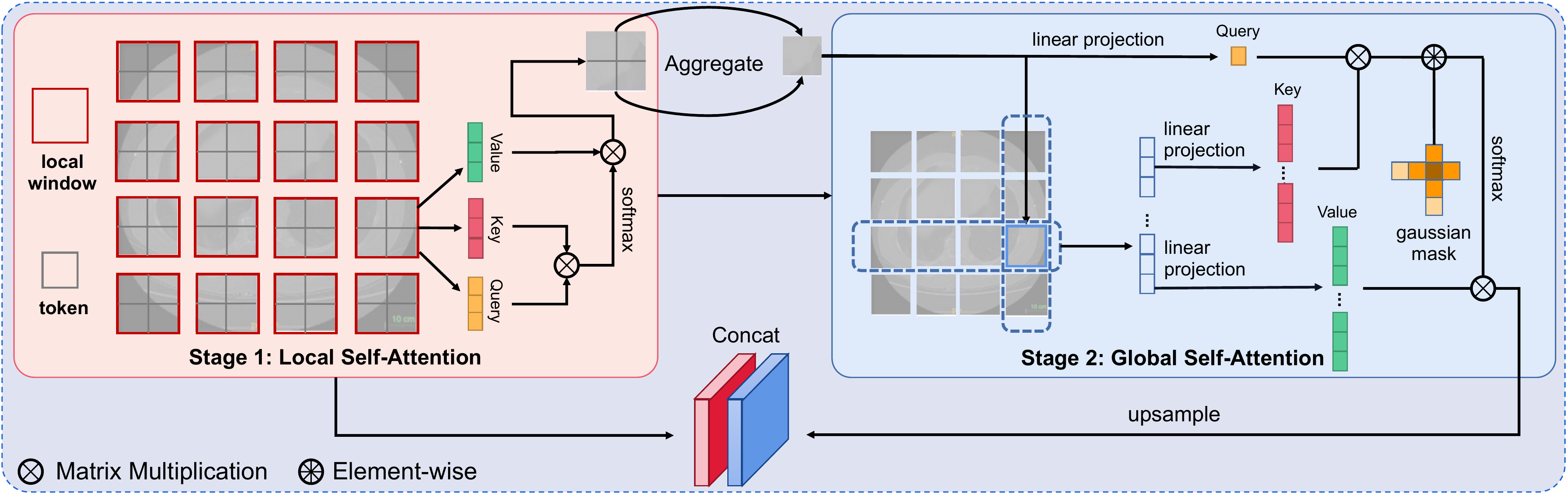}
\caption{Detailed structure of the proposed Local-Global Gaussian-Weighted Self-Attention.}
\label{fig3}
\end{figure*}

LGG-SA perfectly embodies the idea of focalized computation. Unlike traditional SA that pays equal attention to all tokens, LGG-SA can focuses more on nearby regions because of the use of Local-Global strategy and Gaussian mask. Experiments prove that LGG-SA can improve the model performance and save the computational resources. Detailed design of this module is shown in Fig.~\ref{fig3}. 

\subsubsection{Local-Global Self-Attention}
SA aims to capture the inter connections between all the entities of the input sequence. In order to realize the goal, SA introduces three matrices, which are key ($K$), query ($Q$) and value ($V$). The three matrices are linear transforms of the input $X$. However, in computer vision, correlations between nearby areas tends to be more important that those far away, and there is no need to spend equal price for farther areas when computing attention map. Therefore, we propose Local-Global Self-Attention. Local SA calculates self-affinities inside each window. Then, the tokens inside each window are aggregated into a global token, representing the main information of the window. For aggregating functions, we tried stride convolution, Max Pooling and other methods, of which Lightweight Dynamic Convolution (LDConv) \cite{LDconv} performs the best. After having the down-sampled entire feature map, we can then perform Global SA with less expense. Mathematically, for an input feature map $X\in R^{H \times W\times C}$, if we set the window size to $p$ ($p$ is fixed to 4 in our experiment), the overall process can be formulated as:
\begin{equation}
\begin{aligned}
z_{local}=LSA(X),
\end{aligned}
\end{equation}
\begin{equation}
\begin{aligned}
z_{global}=GSA(LDConv(z_{local})),
\end{aligned}
\end{equation}
\begin{equation}
\begin{aligned}
z = Concat(z_{local},Upsample(z_{global})),
\end{aligned}
\end{equation}
where $z$ refers to the module output, $LSA$ represents Local Self-Attention, and $GSA$ denotes the corresponding global operation.

\subsubsection{Gaussian-Weighted Axial Attention}

Unlike $LSA$ using the original SA, we propose Gaussian-Weighted Axial Attention (GWAA) for $GSA$. Inspired by \cite{GaussianTrans}, GWAA enhances each query's perception of nearby tokens through a learnable Gaussian matrix, and meanwhile has a lower time complexity due to Axial Attention \cite{axialattention}. Assuming that $Q\in R^{\frac{H}{p} \times \frac{W}{p}}$ represents the queries obtained from aggregation step, for query $q_{i,j}$ in $Q$, we define $D_{i,j}$ as the Euclidean distances between $q_{i,j}$ and its corresponding $K_{i,j}$ and $V_{i,j}$, where $K_{i,j}$ and $V_{i,j}$ are matrices generated from tokens on $i^{th}$ row and $j^{th}$ column after aggregation. Let the similarity between $q$ and $K$ being $S(q, K)$ and Gaussian weight being $e^{-\frac{D_{i,j}^2}{2\sigma^2}}$, the final output at position $(i, j)$ can be formulated as:
\begin{equation}
\begin{aligned}
z_{i,j}&=e^{-\frac{D_{i,j}^2}{2\sigma^2 }}Softmax(S(q_{i,j},K_{i,j}))V_{i,j}.
\end{aligned}
\end{equation}
Since we want the variance $\sigma$ to be learnable, formula (4) can be equally written as:
\begin{equation}
\begin{aligned}
z_{i,j}=Softmax(-\frac{1}{2\sigma ^2}D_{i,j}^2+S(q_{i,j}, K_{i,j}))V_{i,j},
\end{aligned}
\end{equation}
and we can simply use $w$ to represent the coefficient factor before $D_{i,j}^2$. $wD_{i,j}^2$ also acts as relative position bias, by which we can emphasize the position information in MTM. It improves the model performance for explicitly providing relative relation, which the ordinary absolute positional embedding cannot \cite{tener}.

On the whole, for a given image with $n$ voxels, the time complexity of $LSA$ is $O(n)$ when $p$ is fixed. In contrast, time complexity for $GSA$ is $O(n\sqrt{n})$ due to Axial Attention. Therefore, the overall complexity of our proposed LGG-SA is $O(n\sqrt{n})$.

\subsection{External Attention}
External Attention  (EA) \cite{externalattention} is firstly proposed to solve the problem that SA cannot exploit relations between different samples. Unlike Self-Attention using each sample's own linear transformations to calculate the attention score, in EA, all the samples share two memory units $M_K$ and $M_V$ (as is shown in Fig.~\ref{fig2}), depicting the most essential information of the entire dataset. In our design, an additional linear mapping is used for $Q$ to enlarge its channel, improving the representation learning ability of this module.

Since the time complexity of EA is $O(n)$, the overall time complexity of our MT-UNet stays $O(n\sqrt{n})$.



\begin{table}[t]
\renewcommand\arraystretch{0.883}
\renewcommand\tabcolsep{2.0pt}
\caption{Ablation study on ACDC dataset. '-' stands for not applicable and '$\circ$' denotes incompletely used.}
\label{tab1}
\center
\begin{tabular}{cccccc}
\hline
Method               & LSA & GSA & EA & DSC(\%) & HD95(mm) \\ \hline
ViT Encoder  & -   & -   & -  & 89.38   & 2.54     \\
MTM w/o GSA    & $\checkmark$    & $\times$    & $\checkmark$   & 89.57   & 2.67     \\
MTM w/o LSA    & $\times$    & $\checkmark$    & $\checkmark$   & 89.41   & 4.32     \\
MTM w/o EA       & $\checkmark$    &$\checkmark$     & $\times$   & 89.39   & 3.55     \\
MTM w/o Gaussian & $\checkmark$    & $\circ$    & $\checkmark$   & 89.53   & 2.28     \\
MTM (Ours)       & $\checkmark$    & $\checkmark$    & $\checkmark$   & 90.43   & 2.23     \\ \hline
\end{tabular}
\end{table}

\begin{table*}[t]
\renewcommand\arraystretch{0.883}
\renewcommand\tabcolsep{2.2pt}
\caption{Experimental results of the Synapse Dataset. DSC of each single class is also presented.}
\label{tab2}
\center
\begin{tabular}{ccccccccccc}
\hline
Method       & DSC(\%) & HD95(mm) & Aorta & Gallbladder & Kidney(L) & Kidney(R) & Liver & Pancreas & Spleen & Stomach \\ \hline
V-Net \cite{vnet}       & 68.81   & -        & 75.34 & 51.87       & 77.10     & 80.75     & 87.84 & 40.05    & 80.56  & 56.98   \\
DARR \cite{DARR}        & 69.77   & -        & 74.74 & 53.77       & 72.31     & 73.24     & 94.08 & 54.18    & 89.90  & 45.96   \\
R50 UNet \cite{transunet}    & 74.68   & 36.87    & 84.18 & 62.84       & 79.19     & 71.29     & 93.35 & 48.23    & 84.41  & 73.92   \\
R50 AttnUNet \cite{transunet} & 75.57   & 36.97    & 55.92 & 63.91       & 79.20     & 72.71     & 93.56 & 49.37    & 87.19  & 74.95   \\
UNet \cite{unet}         & 76.85   & 39.70    & 89.07 & 69.72       & 77.77     & 68.60     & 93.43 & 53.98    & 86.67  & 75.58   \\
AttnUNet \cite{attentionunet}    & 77.77   & 36.02    & 89.55 & 68.88       & 77.98     & 71.11     & 93.57 & 58.04    & 87.30  & 75.75   \\ \hline
R50 ViT \cite{transunet}     & 71.29   & 32.87    & 73.73 & 55.13       & 75.80     & 72.20     & 91.51 & 45.99    & 81.99  & 73.95   \\
ViT \cite{ViT}         & 61.50   & 39.61    & 44.38 & 39.59       & 67.46     & 62.94     & 89.21 & 43.14    & 75.45  & 69.78   \\
TransUNet \cite{transunet}    & 77.48   & 31.69    & 87.23 & 63.13       & 81.87     & 77.02     & 94.08 & 55.86    & 85.08  & 75.62   \\
TransClaw U-Net \cite{transclaw}    & 78.09   & 26.38    & 85.87 & 61.38       & 84.83     & 79.36     & 94.28 & 57.65    & 87.74  & 73.55   \\
\textbf{Ours} & \textbf{78.59}   & \textbf{26.59}    & \textbf{87.92} & \textbf{64.99}       & \textbf{81.47}     & \textbf{77.29}     & \textbf{93.06} & \textbf{59.46}    & \textbf{87.75}  & \textbf{76.81} \\ \hline
\end{tabular}
\end{table*}

\begin{table}[t]
\renewcommand\arraystretch{0.883}
\caption{Experimental results of the ACDC Dataset.}
\label{tab3}
\center
\begin{tabular}{ccccc}
\hline
Method       & DSC(\%) & RV    & Myo   & LV    \\ \hline
R50 UNet \cite{transunet}    & 87.60   & 84.62 & 84.52 & 93.68 \\
R50 AttnUNet \cite{transunet} & 86.90   & 83.27 & 84.33 & 93.53 \\ \hline
ViT-CUP \cite{transunet}     & 83.41   & 80.93 & 78.12 & 91.17 \\
R50 ViT  \cite{transunet} & 86.19   & 82.51 & 83.01 & 93.05 \\
TransUNet \cite{transunet}  & 89.71   & 86.67 & 87.27 & 95.18 \\
Swin-Unet \cite{swinunet} & 88.07    & 85.77    &84.42    &94.03 \\
\textbf{Ours}         & \textbf{90.43}        & \textbf{86.64}       & \textbf{89.04}       & \textbf{95.62}      \\ \hline
\end{tabular}
\end{table}

\section{Experiments}
\label{sec:pagestyle}

\subsection{Datasets And Metrics}
\textbf{Synapse.} Synapse is a public multi-organ segmentation dataset. There are 30 contrast-enhanced abdominal clinical CT cases in this dataset. Following the settings in \cite{transunet}, 18 cases are used for training and 12 for testing. The annotation of each image includes 8 abdominal organs (aorta, gallbladder, spleen, left kidney, right kidney, liver, pancreas, spleen and stomach). We use Dice Similarity Coefficient (DSC) and 95\% Hausdorff Distance (HD95) to evaluate our method on this dataset. 

\noindent \textbf{ACDC.} ACDC is a public cardiac MRI dataset consisting of 100 exams. For each exam, there are two different modalities, and the corresponding label includes left ventricle (LV), right ventricle (RV) and myocardium (MYO). Same to the settings of \cite{transunet}, the dataset is split into 70 training samples, 10 validation samples and 20 testing samples.


\subsection{Implementation Details}
All the experiments are conducted on a Nvidia GTX 1080Ti GPU. The input image size is set to $224\times224$ for all the methods. Data augmentation includes random flip and random rotation. All the models are optimized by Adam \cite{adam} with learning rate $1e^{-4}$ and batch size 12. Pre-trained weights are used for other methods if provided, while our model is trained from scratch.

\begin{figure}[t]
\centering
\includegraphics[width=0.48\textwidth]{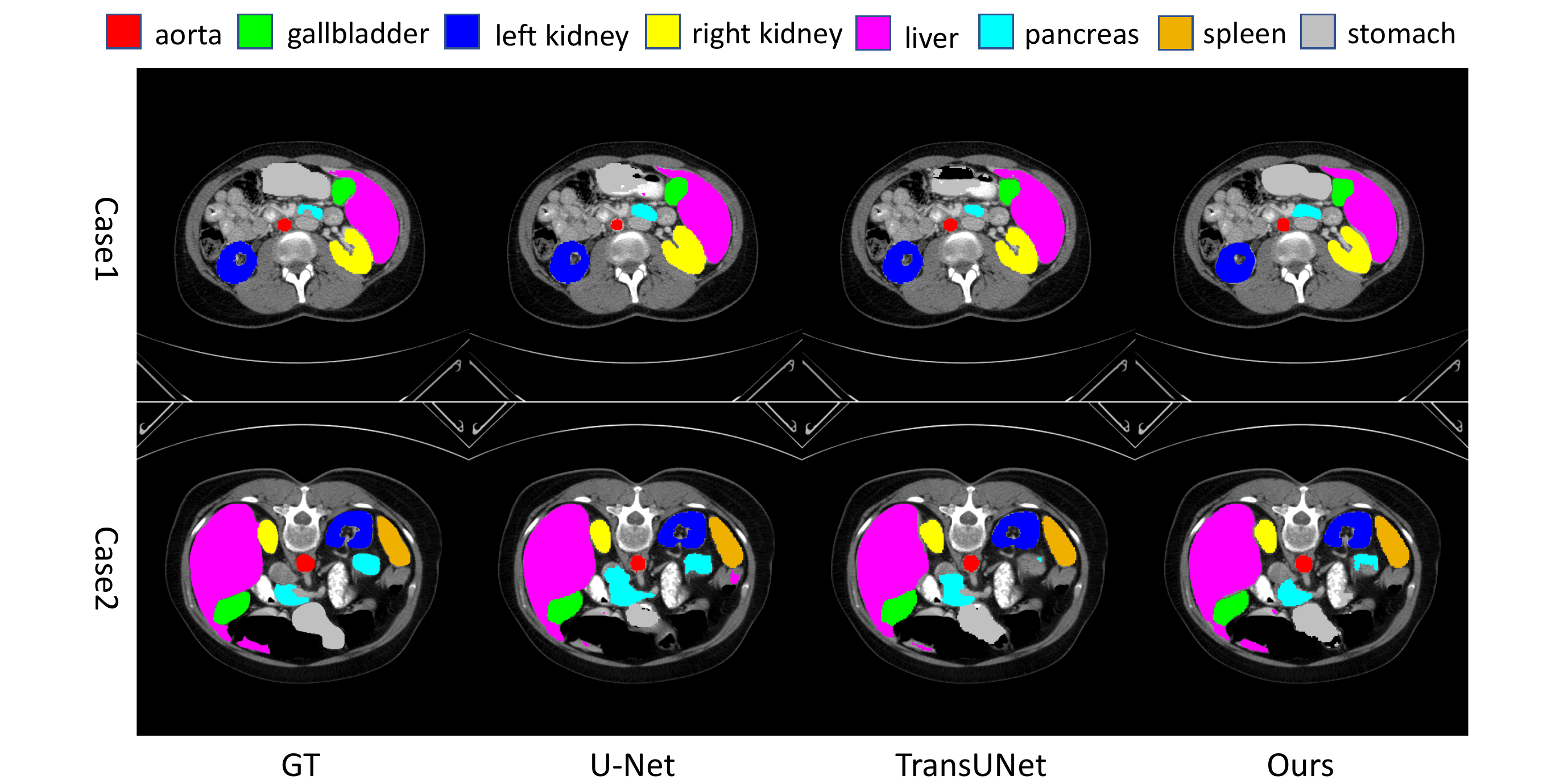}
\caption{Visualization of different methods' segmentation results on Synapse dataset. Best viewed in color.}
\label{fig4}
\end{figure}

\subsection{Ablation Study}
We compared our implementation with other different structures. At first, we tried removing Local SA or Global SA to verify their effectiveness. Then we compared our model with original Transformer. The experimental results are listed in Table.~\ref{tab1}. As is illustrated, neither Local SA nor Global SA is dispensable for the model, since removing any one of them can leads to performance loss. Gaussian mask also proves its necessity for helping the network focus more on local areas. In addition, EA proved its effectiveness as well, since after using it the overall performance gains a 1.04\% and 1.32mm increment in DSC and HD95 respectively. Generally speaking, MTM outperforms original Transformer encoder in the experiment, despite having a even lower time complexity.

\subsection{Experimental Results}
Experimental results on two datasets are presented in Table.~\ref{tab2} and Table.~\ref{tab3} respectively. As is shown, traditional CNNs still have great performance, with Attention-UNet even outperforming TransUNet on Synapse. Nevertheless, on both datasets, our method surpasses CNNs by a large margin, achieving 78.59\% DSC on Synapse and 90.43\% on ACDC. In addition, our method also consistently exceeds Trans-Unet and other Vision Transformers. 

Some segmentation results are presented in Fig.~\ref{fig4}. In Case1, our method shows its overwhelming advantage on segmenting aorta and stomach, which is consistent with the result in Table.~\ref{tab1}. In Case2, our method also surpasses other Vision Transformers in complex shaped organ segmentation (e.g. liver and left kidney) due to its balanced perception for local and global context.

\section{CONCLUSIONS}
\label{sec:typestyle}
In this work, we propose an efficient Vision Transformer named MT-UNet for medical image segmentation. The model is characterized by MTM, which is capable of learning inter- and intra-affinities simultaneously because of LGG-SA and EA. The proposed model has a lower time complexity, and also outperforms other state-of-the-art Vision Transformers in our experiments.

\section{ACKNOWLEDGMENTS}
\label{sec:majhead}
This work was supported in part by Major Scientific Research Project of Zhejiang Lab under the Grant No. 2020ND8AD01, and in part
by the Grant-in Aid for Scientific Research from the Japanese Ministry for Education, Science, Culture and Sports (MEXT) under the Grant No. 20KK0234, No. 21H03470, and No. 20K21821.




\vfill\pagebreak



\bibliographystyle{IEEEbib}
\bibliography{IEEEfull}

\end{document}